\documentstyle[prb,aps,epsfig,rotating,lscape,%
amssymb,amsmath,multicol,float]{revtex}

\begin{document}

\sloppy


\title{Translational Diffusion of Polymer Chains \\
 with Excluded Volume and Hydrodynamic Interactions \\
 by Brownian Dynamics Simulation}
\author{Bo Liu and Burkhard D\"unweg\footnote{Electronic mail: 
       duenweg@mpip-mainz.mpg.de}}
\address{Max--Planck--Institut f\"ur Polymerforschung \\
         Ackermannweg 10, D--55128 Mainz, Germany}

\date{\today}

\maketitle

\widetext

\begin{abstract}
  Within Kirkwood theory, we study the translational diffusion
  coefficient of a single polymer chain in dilute solution, and focus
  on the small difference between the short--time Kirkwood value $D^{(K)}$
  and the asymptotic long--time value $D$. We calculate this
  correction term by highly accurate large--scale Brownian Dynamics
  simulations, and show that it is in perfect agreement with the
  rigorous variational result $D < D^{(K)}$, and with Fixman's
  Green--Kubo formula, which is re--derived. This resolves the puzzle
  posed by earlier numerical results (Rey {\em et al.}, Macromolecules
  24, 4666 (1991)), which rather seemed to indicate $D > D^{(K)}$; the
  older data are shown to have insufficient statistical accuracy to
  resolve this question. We then discuss the Green--Kubo integrand in
  some detail. This function behaves very differently for
  pre--averaged vs. fluctuating hydrodynamics, as shown for the
  initial value by analytical considerations corroborated by numerical
  results. We also present further numerical data on the chain's
  statics and dynamics.
\end{abstract}

\begin{multicols}{2}
\narrowtext


\section{Introduction}

Diffusion of polymer chains in dilute solution has been a subject of
substantial theoretical interest \cite{kirkwood:48,kirkwood:53,%
ikeda:56,zimm:56,erpenbeck:63,fixman:65,fixman:65a,horta:68,%
zimm:80,wilemski:81,doi:86}. A particularly important result is the Kirkwood
formula for the translational diffusion coefficient of a molecule immersed in
a solvent of temperature $T$ and viscosity $\eta$,
\begin{equation} \label{eq:dk0}
D^{(K)} = \frac{D_0}{N} + \frac{k_B T}{6 \pi \eta}
\left\langle \frac{1}{R_H} \right\rangle .
\end{equation}
Here $k_B$ denotes Boltzmann's constant, while $N$ is the number of chain
segments, each of which has a segmental diffusion coefficient $D_0 = k_B T /
\zeta$, where $\zeta$ is the segmental friction coefficient, $\zeta = 6 \pi
\eta a$, where $a$ denotes the segment's Stokes radius. $R_H$ is the
hydrodynamic radius of the molecule, which is defined as a thermal average
over the inverse intramolecular distances between segments:
\begin{equation}
\left\langle \frac{1}{R_H} \right\rangle = 
\frac{1}{N^2} \sum_{i \ne j}
\left\langle \frac{1}{r_{ij}} \right\rangle .
\end{equation}
Equation \ref{eq:dk0} is not exact, but rather the result of a number of
approximations \cite{doi:86}. Firstly, the Brownian motion of the molecule is
described by a Smoluchowski equation (Kirkwood's diffusion equation) in the
space of the segment coordinates $\vec r_i$. Secondly, the diffusion tensor
$\tensor D_{ij}$, whose off--diagonal elements describe the hydrodynamic
interaction between segments $i$ and $j$, is approximated by the Oseen
formula,
\begin{equation}
\tensor D_{ij} = \frac{k_B T}{\zeta} \delta_{ij} \tensor 1 +
(1 - \delta_{ij}) \frac{k_B T}{8 \pi \eta r_{ij}}
\left( {\tensor 1} + \hat{r}_{ij} \otimes \hat{r}_{ij} \right) ,
\end{equation}
where $\hat{r}_{ij}$ is the unit vector in direction of $\vec r_{ij} = \vec
r_i - \vec r_j$. Thirdly, the diffusion coefficient is evaluated in the {\em
  short--time limit} of the Smoluchowski equation. The diffusion coefficient
in the asymptotic long--time limit, which we denote as $D$, is {\em different}
from the short--time value $D^{(K)}$, due to the presence of intramolecular
dynamic correlations, which are expected to decay on the time scale of the
overall relaxation time of the chain's conformational degrees of freedom (the
Zimm time $\tau_Z$). Thus, the behavior on short time scales $t \ll \tau_Z$ is
governed by $D^{(K)}$, while on long time scales $t \gg \tau_Z$ the diffusive
motion takes place with the coefficient $D$. For a general diffusion tensor
(not necessarily of the Oseen type) the same considerations apply. In this
case the short--time diffusion coefficient is obtained by averaging over the
trace:
\begin{equation} \label{eq:dk}
D^{(K)} = \frac{1}{3 N^2} \sum_{ij} \mbox{\rm Tr} 
\left\langle \tensor D_{ij} \right\rangle ;
\end{equation}
in what follows, the term ``Kirkwood formula'' will always refer to Eq.
\ref{eq:dk}.

The difference between $D^{(K)}$ and $D$ has caused considerable interest and
some confusion (see below); it is the purpose of the present paper to
contribute to the clarification of these questions. Both light scattering
experiments \cite{akcasu:80,han:81,tsunashima:87,bhatt:89} and
first--principles Molecular Dynamics (MD) simulations
\cite{pierleoni:92,duenweg:93} show that the difference must be quite small
such that it is below resolution. In the case of MD, the main difficulty is
the fact that the ``short--time'' limit of Kirkwood theory is of course not the
exact $t \to 0$ limit, but rather a time regime where the local motions of the
particles in their ``cages'' (or the momenta) have already relaxed completely,
while the conformational degrees of freedom of the chain must at the same time
be considered as completely unrelaxed. To reach this separation of time scales
(i.~e. ballistic~/ short--time diffusive~/ long--time diffusive) in an
unambiguous fashion would require extremely long chains which are not
accessible to today's computer power. A related problem is the fact that the
monomer diffusion coefficient $D_0$, which must be taken into account in order
to evaluate $D^{(K)}$ correctly \cite{duenweg:93}, is a somewhat fuzzy and
ill--defined object in MD. This is seen in a quite obvious fashion when
looking at the velocity autocorrelation function of a single monomer, whose
time integral is the chain diffusion constant $D$ \cite{hansen:86}: After a
rather quick short--time decay, one gets a negative tail which slowly decays
to zero on the time scale of the overall relaxation of the chain as a whole
\cite{kopf:97}. This latter decay describes the Rouse-- or Zimm--like slowing
down by the coupling to the other monomers, while the quick short--time regime
defines, in principle, the monomer diffusion coefficient $D_0$. However, there
is no obvious and well--defined time which discriminates between short-- and
long--time regime. Depending on where this ``cut'' is being put, one obtains
slightly different values for $D_0$. This latter problem does not exist in a
recently introduced hybrid approach \cite{ahlrichs:99}, where the chain
monomers are coupled dissipatively to a discretized hydrodynamic background,
such that the monomer diffusion coefficient is an input parameter. However,
the former problem (lack of separation of time scales) is still present.

In order to cope with these difficulties, one therefore has to work {\em
  strictly within} the framework of Kirkwood theory (i.~e. the Smoluchowski
equation), such that $D_0$ is just an input parameter, the ballistic regime
does not exist, and the theory has a well--defined diffusive short--time
limit. Nevertheless, calculating the long--time diffusion coefficient $D$, and
the smooth crossover between the short--time behavior governed by $D^{(K)}$,
and the long--time regime governed by $D$, requires to actually solve the
Smoluchowski equation. However, closed analytical solutions do not exist as
soon as fluctuating hydrodynamic interactions and~/ or excluded volume
interactions are introduced. For this reason, analytical progress on the
problem has been limited; nevertheless, a few important results have been
obtained. Within the preaveraged approximation, \"Ottinger \cite{oettinger:87}
derived a closed expression for $D$ which clearly differs from $D^{(K)}$.
Furthermore, it is possible to establish rigorous variational bounds on
transport coefficients \cite{doi:86,rotne:69,fixman:83}, from which one can
show that $D$ must be {\em smaller} than $D^{(K)}$. Moreover, Fixman
\cite{fixman:81} derived the Green--Kubo type relation
\begin{equation}
D = D^{(K)} - D_1
\end{equation}
with
\begin{equation} \label{eq:fixgreenkubo}
D_1 = \frac{1}{3 N^2} \int_0^\infty dt
      \left< \vec A (0) \cdot \vec A (t) \right> > 0
\end{equation}
and
\begin{equation}
\vec A = \sum_{ij} \tensor \mu_{ij} \cdot \vec F_j ,
\end{equation}
where $\vec F_j$ is the force acting on segment $j$, and
$\tensor \mu_{ij}$ is the mobility tensor, $\tensor \mu_{ij}
= \tensor D_{ij} / (k_B T)$. 

However, for a more quantitative understanding one has to rely on
numerical solutions of the Smoluchowski equation. The most common
approach is Brownian Dynamics (BD) simulations, which, for the case of
fluctuating hydrodynamic interactions, was pioneered by Ermak and
McCammon \cite{ermak:78}. BD simulations for polymer chains have been
done by numerous workers before
\cite{fixman:81,fixman:83a,fixman:86,lewis:88,%
zylka:89,rey:89,rey:91,rey:92,oettinger:94m,oettinger:94p,%
lyulin:99,jendrejack:00}. One of these studies \cite{rey:91} produced
quite accurate data, which however seemed to indicate $D > D^{(K)}$.
No satisfactory explanation of this deviation has been given so
far. For this reason, we have looked into the problem again, and
performed careful BD simulations of the same model with identical
parameters. In this model, excluded volume interactions are
represented by a relatively soft potential, while the hydrodynamic
interactions are modeled via the Rotne--Prager--Yamakawa (RPY) tensor
\cite{rotne:69,yamakawa:70}.

As noted before \cite{rey:91}, this is a numerically demanding task,
as such a calculation is looking for a rather small effect. One has,
on the one hand, to watch out for discretization errors due to the
finite time step \cite{rey:91}, and, on the other hand, to make sure
that the statistical accuracy of the data suffices. In order to test
the relation $D = D^{(K)} - D_1$, all three quantities must be
determined independently. As our analysis shows (see below), it turns
out that both $D^{(K)}$ and $D_1$ can be calculated with moderate
statistical effort, and Ref. \onlinecite{rey:91} has indeed obtained
values which coincide with ours. The diffusion coefficient $D$ itself,
however, is subject to much larger fluctuations, such that its
accurate determination needs substantially more computer power than
what was available to the authors of Ref. \onlinecite{rey:91}. We
believe that this is the most likely explanation of the deviation,
which is not confirmed by our much more accurate data (these are based
on an observation time which is orders of magnitude larger than that
of Ref. \onlinecite{rey:91}). Rather, we find a very nice agreement
with the relation $D = D^{(K)} - D_1$ with $D_1 > 0$. We also
re--derive Fixman's \cite{fixman:81} Green--Kubo formula for $D_1$,
Eq. \ref{eq:fixgreenkubo}, by means of an alternative approach based
on the Mori--Zwanzig projection operator technique, and try to
elucidate its relation to the stochastic analog of the velocity
autocorrelation function. Furthermore, we attempt to discuss the
scaling behavior of $D_1$ via both scaling arguments and numerical
data; here we have obtained some results which we view as somewhat
unexpected and counter--intuitive. However, these should be considered
as preliminary and inconclusive. The computational demands of the
calculations are so large that it is impossible to accurately study
the dynamics for chain lengths much beyond $N \sim 10^2$, which is
clearly not long enough to justify any claim of asymptotic behavior.

The outline of this article is as follows. In Sec. \ref{sec:method} we define
the simulated model and describe the simulation procedure we applied. In Sec.
\ref{sec:diffusion} we derive the short--time diffusion coefficient and
long--time diffusion coefficient from the mean square displacement of the
center of mass, and compare the analytical predictions with the numerical
results. Section \ref{sec:scalinga2} presents our considerations concerning
the scaling of the correction term $D_1$ together with numerical data. Section
\ref{sec:results} then discusses further data on the static and dynamic
scaling properties, mainly to demonstrate the consistency of the results. In
Sec. \ref{sec:summary} we conclude with some final remarks.

\section{Model and Simulation Method}
\label{sec:method}

\subsection{Bead--Spring Model}

Following Rey {\em et al.} \cite{rey:91}, we have modeled the flexible polymer
as a linear chain constituted by $N$ units ranging from $N=6$ to $200$, each one
connected with its first neighbors by means of harmonic springs. Therefore the
bond lengths follow a Gaussian distribution whose variance we denote by $b^2$,
i.~e. $b$ is the statistical segment length. Excluded volume forces only act
between non--neighboring units; they are introduced by means of a potential of
the form $A \exp (- \beta r_{ij})$, where $A$ and $\beta$ are constant
parameters. This potential is cut off at a distance $r_{c}$.

\subsection{Hydrodynamic Interaction}

The hydrodynamic interaction is introduced through the diffusion tensor
proposed by Rotne and Prager and Yamakawa (RPY) \cite{rotne:69,yamakawa:70},
\begin{equation}\label{eq:HI}
\tensor D_{ij} =
\left\{
\begin{array}{l l}
\frac{k_{B}T}{6\pi \eta a} \tensor 1  & \text{$i=j$} \\
\frac{k_{B}T}{8\pi \eta r_{ij}}
    [ \left( \tensor 1 + \hat{r}_{ij} \otimes \hat{r}_{ij} \right) \\
    \hspace{0.25cm}
    + \frac{2a^{2}}{r^{2}_{ij}} \left( \frac{1}{3} \tensor 1
    - \hat{r}_{ij} \otimes \hat{r}_{ij} \right) ], 
    &\text{$i\neq j$ and $r_{ij}\geqslant 2a$}\\
\frac{k_{B}T}{6\pi \eta a}
    [ \left(1-\frac{9}{32} \frac{r_{ij}}{a}\right) \tensor 1 \\
    \hspace{0.25cm}
    +\frac{3}{32} \frac{{r}_{ij}}{a} \hat{r}_{ij} \otimes \hat{r}_{ij} ],
    &\text{$i\neq j$ and $r_{ij}<2a$}
\end{array} \right.
\end{equation}
where $a$ is the Stokes radius of the beads. This tensor is positive--definite
for all chain configurations.

\subsection{Algorithm}

The polymer motion is governed by the stochastic differential equation that we
solve through the first--order Ermak and McCammon algorithm \cite{ermak:78},
\begin{equation}\label{eq:ermak}
\vec{r}_{i} (t+h) = \vec{r}_{i}(t) 
+ \sum _{j} \tensor \mu_{ij} \cdot \vec{F}_{j} h + \vec{\rho_{i}} ,
\end{equation}
where $h$ is the time step, $\vec F_{i}$ the force on monomer $i$, $\tensor
\mu_{ij} = \tensor D_{ij} / (k_B T)$, and $\vec \rho_{i}$ a random
displacement with zero mean and variance--covariance matrix given by
\begin{equation} \label{eq:covariance}
\left\langle \vec{\rho _{i}} \otimes \vec{\rho_{j}} \right\rangle
= 2 \tensor D_{ij} h.
\end{equation}
We generated the random terms $\vec \rho_i$ from a uniform distribution
\cite{nongauss}, and used the procedure given in Ref. \onlinecite{ermak:78} to
satisfy Eq. \ref{eq:covariance}. The term $h \sum_{j} \left( \partial /
  \partial \vec r_j \right) \cdot \tensor D_{ij}$ on the right hand side of
Eq.  (\ref{eq:ermak}) was omitted, because the RPY tensor, due to the
incompressibility of the solvent, is divergence--free, $\left( \partial /
  \partial \vec r_j \right) \cdot \tensor D_{ij} = 0$.

\subsection{Simulation Details}

As in Ref. \onlinecite{rey:91}, we use a unit system where the three
quantitities $b$, $k_B T$, and $\zeta = 6 \pi \eta a$ are set to
unity. Hence the time unit is given by $\zeta b^{2} / k_{B} T$. In
this unit system, the other parameters are $A = 75$, $\beta = 4$,
$r_{c} = 0.512$, and $a=0.256$. We used a reduced time step of
$h=0.005$, which is twice as small as that of
Ref. \onlinecite{rey:91}. This choice was motivated by a comparison of
end--to--end distance and gyration radius data obtained by simulations
with various $h$ with the corresponding data from Monte Carlo
runs. This comparison showed that highly accurate results require a
rather conservative $h$ value.

For $N \le 50$, data with good statistics were generated by the
following procedure: (i) Equilibration by BD without hydrodynamic
interactions, (ii) short additional equilibration with hydrodynamic
interactions turned on, and (iii) a long BD production run.
Furthermore, we averaged over five independent such runs. For these
chains, the resulting total simulation time divided by the longest
relaxation time $\tau_{Z}$ (which can be obtained from the time
autocorrelation function of the end--to--end distance) is $0.5
\times 10^6$ for $N \le 35$, $0.4 \times 10^6$ for $N = 40$,
$0.3 \times 10^6$ for $N = 45$, and $0.2 \times 10^6$ for
$N = 50$. All the statistical error bars were
estimated by the blocking method developed by Flyv\-bjerg and
Petersen\cite{flyvbjerg:89}.

{From} the results for these chains, it turned out that the dynamics
on short time scales significantly below $\tau_Z$ is of particular
interest (see below). For this reason, we studied two longer chains
($N = 100$ and $200$) by a somewhat different procedure: By an
efficient Monte Carlo procedure (the pivot algorithm
\cite{lal:69,stellman:72} combined with local moves) we generated a
large sample of statistically independent conformations ($128$
conformations for $N = 100$, $3700$ conformations for $N = 200$), from
which we started BD runs with hydrodynamic interactions of short to
moderate length (in our time units: $\tau = 2.5 \times 10^4$ for $N =
100$, $\tau = 100$ for $N = 200$). Although this implies a quite good
statistical accuracy, it is nevertheless significantly worse than for
the shorter chains. Therefore only the most important and interesting
properties were evaluated for $N = 100$ and $200$.

For chain length $N = 6$, our program needed 60 sec. to run
$10^5$ BD steps on a 667 MHz PC. This number of course
strongly increases with $N$ ($\sim 500$ sec. for $N = 50$,
$\sim 10^5$ sec. for $N = 200$);
the asymptotic $N^3$ scaling of the algorithm was observed
for (roughly) $N \ge 30$. The calculations ran for several
months on a 16--machine PC cluster; the overall effort
of the project in terms of single--processor time is
estimated as roughly 6.7 years.

\section{Translational Diffusion Coefficient}
\label{sec:diffusion}

\subsection{Theory}

In order to calculate the translational diffusion coefficient in the short--
and long--time regimes, we study the mean square discplacement of the center
of mass,
\begin{equation}
\vec R_{CM} = \frac{1}{N} \sum_i \vec r_i .
\end{equation}
{From} the Ermak--McCammon algorithm, Eq. \ref{eq:ermak}, we find the updating
rule for $\vec R_{CM}$ in one time step:
\begin{eqnarray}
\Delta \vec R (t) & \equiv & \nonumber
\vec R_{CM} (t + h) - \vec R_{CM} (t) \\
& = & \frac{1}{N} \left( h \vec A (t) + h^{1/2} \vec B (t) \right) ,
\end{eqnarray}
where we have introduced the abbreviations
\begin{equation}
\vec A = \sum_{ij} \tensor \mu_{ij} \cdot \vec F_j
\end{equation}
(this definition is identical to the notation of Fixman \cite{fixman:81}) and
\begin{equation}
\vec B = h^{-1/2} \sum_i \vec \rho_i .
\end{equation}
Note that $\vec A$ and $\vec B$ are defined in such a way that they are
independent of the time step $h$. As $\vec A$ and $\vec B$, evaluated
at the same time, are uncorrelated, and since $\left< \vec B \right>$
vanishes, we find
\begin{eqnarray}
\left< \left( \Delta \vec R \right)^2 \right> 
& = & \nonumber
\frac{h}{N^2} \left< \vec B^2 \right>
+ \frac{h^2}{N^2} \left< \vec A^2 \right> \\
& = &
\frac{2 h}{N^2} \sum_{ij} \mbox{\rm Tr}
\left< \tensor D_{ij} \right> + O(h^2) .
\end{eqnarray}
Thus the short--time diffusion coefficient is just given by the Kirkwood
formula:
\begin{eqnarray}
D^{short} & = & \nonumber
\lim_{h \to 0} \frac{1}{6h} 
\left< \left( \Delta \vec R \right)^2 \right> \\
& = & \frac{1}{3 N^2} \sum_{ij} \mbox{\rm Tr}
\left< \tensor D_{ij} \right>
\equiv D^{(K)} .
\end{eqnarray}

For longer times ($n$ time steps) we evaluate the mean square displacement as
\begin{eqnarray}
& & 
\left< \left( \vec R_{CM} (n h) - \vec R_{CM} (0) \right)^2 \right> \\
& = & \nonumber
\left< \left( \sum_{k = 0}^{n - 1} \Delta \vec R (k h) \right)^2 \right> 
=  
\sum_{kl} \left< \Delta \vec R (k h) \cdot \Delta \vec R(l h) \right> .
\end{eqnarray}
The matrix with elements $\left< \Delta \vec R (k h) \cdot \Delta \vec R(l h)
\right>$ is obviously symmetric. Since all elements with constant $k - l$ are
identical, for reasons of translational symmetry in time, we can simplify the
previous expression as
\begin{eqnarray} \label{eq:exploittimesymmetry}
&& \left< \left( \vec R_{CM} (n h) - \vec R_{CM} (0) \right)^2 \right> \\
& = & \nonumber
n \left< \Delta \vec R (0)^2 \right> +
2 \sum_{k = 1}^{n - 1} (n - k) 
\left< \Delta \vec R (0) \cdot \Delta \vec R (k h) \right> .
\end{eqnarray}
This is quite analogous to the standard relation between mean square
displacement and velocity autocorrelation function \cite{hansen:86}. In
the long--time limit $n \to \infty$ we thus obtain a diffusion coefficient
which still depends on the time step:
\begin{eqnarray} \label{eq:longtimelimit}
D^{long} (h) & = & \nonumber
\lim_{n \to \infty} \frac{1}{6 n h}
\left< \left( \vec R_{CM} (n h) - \vec R_{CM} (0) \right)^2 \right> \\
& = & 
D^{(K)} + \frac{h}{6 N^2} \left< \vec A^2 \right> \\
& + & \nonumber
\frac{1}{3 h} \sum_{k = 1}^\infty 
\left< \Delta \vec R (0) \cdot \Delta \vec R (k h) \right> ;
\end{eqnarray}
here we have assumed, as usual, that the correlation function
decays quickly to zero.

\begin{figure}
\begin{center}
\epsfig{file=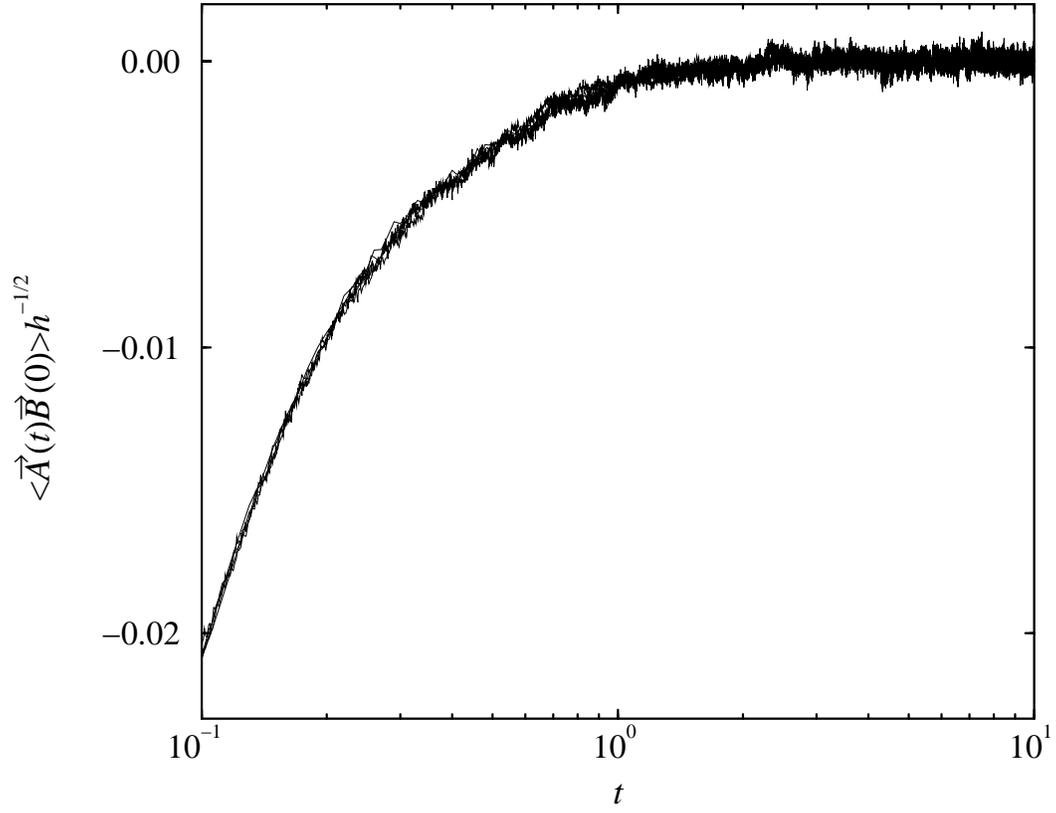,width=11cm,angle=-90}
\end{center}
\caption{The correlation function $\left\langle \vec
  A(t) \cdot \vec B(0)\right\rangle$, scaled with $h^{-1/2}$,
  for $N = 6$ and various time steps $h$ = 0.0005, 0.001,
  0.002, 0.005 and 0.01, respectively.}
\label{fig:deltd2}
\end{figure}

We now make use of the fact that the stochastic terms $\vec B$ are
uncorrelated at different times, and that there is a correlation
between $\vec A$ and $\vec B$ only if $\vec A$ is evaluated at
a {\em later} time as $\vec B$ (of course, a stochastic ``kick''
at a certein time will influence how the system evolves dynamically
in the future). We thus obtain for $k \ge 1$
\begin{eqnarray}
\left< \Delta \vec R (0) \cdot \Delta \vec R (k h) \right>
& = & \nonumber
\frac{h^2}{N^2} \left< \vec A (0) \cdot \vec A (k h) \right> \\
& + &
\frac{h^{3/2}}{N^2} \left< \vec B(0) \cdot \vec A (k h) \right> ,
\end{eqnarray}
yielding the relation
\begin{equation}
D^{long} (h)  = D^{(K)} + D_1 (h) + D_2 (h)
\end{equation}
with
\begin{equation}
D_1 (h) = \frac{h}{6 N^2} \left< \vec A^2 \right>
        + \frac{h}{3 N^2} \sum_{k = 1}^\infty 
          \left< \vec A (0) \cdot \vec A (k h) \right>
\end{equation}
and
\begin{equation}
D_2 (h) = \frac{h^{1/2}}{3 N^2} \sum_{k = 1}^\infty
          \left< \vec B(0) \cdot \vec A (k h) \right> .
\end{equation}
In the continuum limit $h \to 0$, we obviously have
\begin{equation} \label{eq:defined1}
D_1 = \frac{1}{3 N^2} \int_0^\infty dt
      \left< \vec A (0) \cdot \vec A (t) \right> ,
\end{equation}
where the previous formulae tell us how the integral should be consistently
discretized. Concerning $D_2$, one obtains
\begin{equation}\label{eq:defined2}
D_2 = \frac{1}{3 N^2 h^{1/2}} \int_0^\infty dt 
      \left< \vec B (0) \cdot \vec A (t) \right> .
\end{equation}
At first glance, this looks as if this contribution would diverge for $h \to
0$, but this is not the case. Rather, the correlation function $\left< \vec B
  (0) \cdot \vec A (t) \right>$ depends on the time step, and is, to leading
order, proportional to $h^{1/2}$, such that $D_2$ converges to a well--defined
non--trivial value. This is demonstrated in Fig. \ref{fig:deltd2}, where
$\left< \vec B (0) \cdot \vec A (t) \right> h^{-1/2}$ is plotted for $N = 6$
and various time steps $h$. The $h^{1/2}$ dependence may be explained by
linear response theory: As $\left< \vec A \right>$ vanishes for symmetry
reasons, only that part of $\vec A$ will contribute to $\left< \vec B (0)
  \cdot \vec A (t) \right>$ which is actually the response to the ``kick'' at
time zero.  This ``kick'', however, has an infinitesimally small amplitude of
order $h^{1/2}$. Therefore, linear response theory should be applicable, and
the response in $\vec A$ should be proportional to $h^{1/2}$ as well.

On a more formal level, we can write the Ermak--McCammon updating rule (Eq.
\ref{eq:ermak}) in the continuum limit as a Langevin equation with Ito
interpretation:
\begin{equation}
\frac{d}{dt} \vec r_i = \sum_j \tensor \mu_{ij} \cdot \vec F_j + \vec f_i ,
\end{equation}
where the noise term $\vec f_i$ has zero mean, and variance
\begin{equation}
\left< \vec f_i (t) \otimes \vec f_j (t^\prime) \right> = 
2 \tensor D_{ij} \delta (t - t^\prime) ,
\end{equation}
i.~e. $\vec f_i$ corresponds to $\vec \rho_i / h$. This in turn implies $\vec
B = h^{1/2} \sum_i \vec f_i$, such that $D_2$ can also be written as
\begin{equation} \label{eq:d2finalresult}
D_2 = \frac{1}{3 N^2} \sum_i \int_0^\infty dt 
      \left< \vec f_i (0) \cdot \vec A (t) \right> .
\end{equation}
In seeming contrast to this result ($D = D^{(K)} + D_1 + D_2$), Fixman
\cite{fixman:81} rather obtained from linear response theory $D = D^{(K)} -
D_1$, where $D_1$ is defined precisely as in Eq. \ref{eq:defined1}. His result
is however as valid as ours, and in what follows we will give an alternative
derivation based on the Mori--Zwanzig projection operator formalism
\cite{forster:86}. The only conclusion is that $D_1$ and $D_2$ must satisfy
the relation $D_2 = - 2 D_1$. Unfortunately, we have not been able to derive
this result directly; however, it is very nicely borne out by our numerical
data (Table \ref{table:d}). Furthermore, if $\left< \vec A (0) \cdot \vec A
  (t) \right>$ is positive for all times (as it is the case for our simulation
data), it is immediately obvious that $D$ must indeed be smaller than
$D^{(K)}$.

For the Mori--Zwanzig analysis of $D$, we first notice that the Langevin
equation corresponds to the Fokker--Planck equation (Kirkwood diffusion
equation)
\begin{equation}
\frac{\partial}{\partial t} P \left( \Gamma, t \vert \Gamma^0, 0 \right)
= - i {\cal L} P \left( \Gamma, t \vert \Gamma^0, 0 \right) ,
\end{equation}
where $\Gamma$ is a shorthand notation for the set of all monomer coordinates,
$P \left( \Gamma, t \vert \Gamma^0, 0 \right)$ is the transition probability
density for the system going from $\Gamma^0$ at time $0$ to $\Gamma$ at time
$t$, and $- i {\cal L}$ is the Fokker--Planck operator
\begin{equation}
- i {\cal L} = \sum_{ij} \frac{\partial}{\partial \vec r_i} \cdot
\tensor D_{ij} \cdot \left( \frac{\partial}{\partial \vec r_j}
- \beta \vec F_j \right) ,
\end{equation}
where $\beta = 1 / (k_B T)$. The formal solution is $P = \exp(- i {\cal L} t)
\delta(\Gamma - \Gamma^0)$, while the equilibrium distribution (i.~e. the
$t \to \infty$ solution) is
\begin{equation}
\rho (\Gamma) = \frac{ \exp (- \beta U) }{\int d\Gamma \exp (- \beta U)} ,
\end{equation}
where $U$ is the potential energy, such that $\vec F_i = - \partial U /
\partial \vec r_i$, and
\begin{equation} \label{eq:rhocommutator}
\rho \frac{\partial}{\partial \vec r_i} f =
\left( \frac{\partial}{\partial \vec r_i} - \beta \vec F_i \right) \rho f
\end{equation}
for an arbitrary function $f(\Gamma)$. Except for $- i {\cal L}$, we also
need $i {\cal L}^\dagger$, which is the adjoint operator of $ - i {\cal L}$
with respect to the standard scalar product $(f \vert g) = \int d\Gamma
f(\Gamma)^\star g(\Gamma)$ ($f^\star$ denoting the complex conjugate),
\begin{equation}
i {\cal L}^\dagger = \sum_{ij} \left( \frac{\partial}{\partial \vec r_j}
+ \beta \vec F_j \right) \cdot \tensor D_{ij} \cdot 
\frac{\partial}{\partial \vec r_i} ,
\end{equation}
as well as $- i \hat{\cal L}$, which is the adjoint operator of $i {\cal
  L}^\dagger$ with respect to the natural scalar product
\begin{equation}
\left< f \vert g \right> = \int d\Gamma \rho(\Gamma)
f(\Gamma)^\star g(\Gamma) .
\end{equation}
{From} Eq. \ref{eq:rhocommutator}, and partial integration, one finds that $- i
\hat{\cal L}$ coincides with $i {\cal L}^\dagger$.

We now use the standard memory equation as derived in Ref.
\onlinecite{forster:86}. In Ref. \onlinecite{duenweg:93a} it was shown how to
generalize this to the case of non--Hamiltonian dynamics; specifically it was
shown there that the integral over the time correlation function of a slow
variable $S$ satisfies the relation
\begin{eqnarray} \label{eq:memory}
&& \int_0^\infty dt \left< S(0)^\star S(t) \right> =
   - \left< S \vert S \right>^2 \times  \\
&& \left[ \left< S \vert i {\cal L}^\dagger \vert S \right> +
\int_0^\infty dt \left< S \vert i {\cal L}^\dagger {\cal Q}
\exp ( i {\cal L}^\dagger t) {\cal Q} i {\cal L}^\dagger \vert S \right>
\right]^{-1} . \nonumber
\end{eqnarray}
Here ${\cal Q}$ is the operator which projects onto the orthogonal space of
the slow variable, i.~e. onto the space of variables which are (statically)
uncorrelated with $S$. For the analysis of diffusion we study the variable
\begin{equation}
S = \exp ( i \vec q \cdot \vec R_{CM} )
\end{equation}
in the limit $q \to 0$ such that $q^{-1}$ is much larger than the polymer
gyration radius. Therefore,
\begin{equation}
\left< S(0)^\star S(t) \right> =
\left< \exp \left[ i \vec q \cdot 
( \vec R_{CM} (t) - \vec R_{CM} (0) ) \right] \right> .
\end{equation}
For times of order of the Zimm relaxation time, or smaller, this
correlation function is very close to unity, due to the smallness
of $q$. For times much larger, the motion of $\vec R_{CM}$ is just
a Gaussian random walk with diffusion constant $D$, and hence
\begin{eqnarray}
\left< S(0)^\star S(t) \right> & = & \nonumber
\exp \left( - \frac{q^2}{6} \left< (\vec R_{CM} (t) - \vec R_{CM} (0))^2
\right> \right) \\
& = & \exp (- D q^2 t ) .
\end{eqnarray}
Thus the left hand side of Eq. \ref{eq:memory} is just
\begin{eqnarray}
\int_0^\infty dt \left< S(0)^\star S(t) \right> = \frac{1}{D q^2} .
\end{eqnarray}
As $\left< S \vert S \right> = 1$, we have
\begin{eqnarray} \label{eq:diffusion_memory}
D & = & - q^{-2} \left< S \vert i {\cal L}^\dagger \vert S \right>
\nonumber \\
& & - q^{-2}
\int_0^\infty dt \left< S \vert i {\cal L}^\dagger {\cal Q}
\exp ( i {\cal L}^\dagger t) {\cal Q} i {\cal L}^\dagger \vert S \right>
\end{eqnarray}
in the limit $q \to 0$. Now, straightforward evaluation yields
\begin{eqnarray}
i {\cal L}^\dagger S & = & \sum_{ij} \left( \frac{i \vec q}{N}
+ \beta \vec F_j \right) \cdot \tensor D_{ij} 
\cdot \frac{i \vec q}{N} \, S
\nonumber \\
& = & - \frac{q^2}{N^2} \sum_{ij} \hat{q} \cdot
\tensor D_{ij} \cdot \hat{q} \, S
+ \frac{i q}{N} \hat{q} \cdot \vec A \, S
\end{eqnarray}
and, for $q \to 0$,
\begin{equation}
\left< S \vert i {\cal L}^\dagger \vert S \right> =
- \frac{q^2}{N^2} \sum_{ij} \frac{1}{3} \mbox{\rm Tr}
\left< \tensor D_{ij} \right> 
= - q^2 D^{(K)}
\end{equation}
(the $\vec A$ term vanishes upon averaging, for symmetry reasons). This also
means that in the limit $q \to 0$ the variable $i {\cal L}^\dagger S$ becomes
orthogonal to $S$, implying that in this limit we can ignore the operator
${\cal Q}$ in Eq. \ref{eq:diffusion_memory}. Furthermore, in the memory
integral it is sufficient to just take the term linear in $q$ for $i {\cal
  L}^\dagger S$ --- any higher order would not contribute to $D$ in the limit
$q \to 0$. In this order we find $i {\cal L}^\dagger S = i \vec q \cdot \vec A
/ N$. As $- i \hat{\cal L} = i {\cal L}^\dagger$, the memory term becomes
\begin{eqnarray}
&& \nonumber
\int_0^\infty dt \left< S \vert i {\cal L}^\dagger {\cal Q}
\exp ( i {\cal L}^\dagger t) {\cal Q} i {\cal L}^\dagger \vert S \right> \\
& = & 
\frac{q^2}{3 N^2}
\int_0^\infty dt \left< \vec A (0) \cdot \vec A (t) \right> 
= q^2 D_1 .
\end{eqnarray}
Combining these results, we obtain Fixman's \cite{fixman:81} formula
$D = D^{(K)} - D_1$.

\subsection{Numerical Results}

\begin{figure}
 \begin{center}
 \epsfig{file=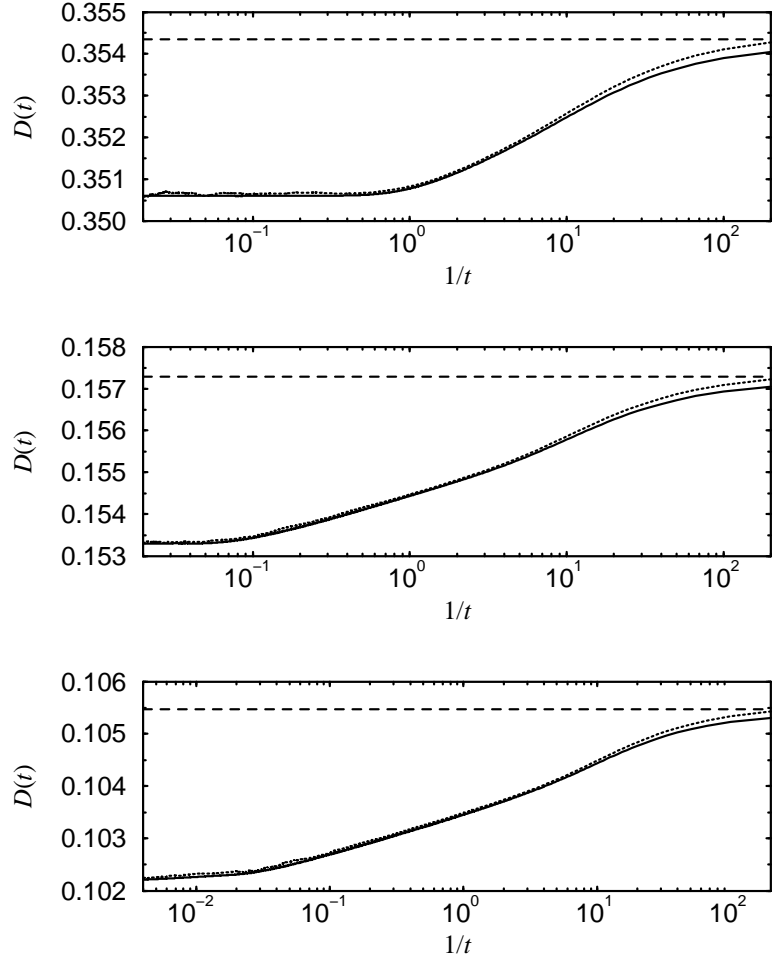,width=13cm,angle=-90}
 \end{center}
\caption{Time--dependent diffusion coefficients $D(t) = D^{(K)} + D_1
  (t) + D_2(t)$ (dotted lines) and $\tilde D(t) = D^{(K)} - D_1 (t)$
  (full lines; see also Eq. \ref{eq:definedoft} etc.), for $N = 6$
  (upper graph), $N = 25$ (middle graph) and $N = 50$ (lower
  graph). The dashed lines indicate the corresponding Kirkwood values
  $D^{(K)}$. }
\label{fig:n6n25n50}
\end{figure}

As we have seen in the previous subsection, the relation $2 D_1 + D_2
= 0$ should hold. Our data (see Table \ref{table:d}) indeed confirm
this prediction. Interestingly enough, we have found numerically that
even the Green--Kubo integrands satisfy the corresponding relation
\begin{equation}
\Delta (t) \equiv 
2 \left< \vec A (0) \cdot \vec A (t) \right> +
h^{-1/2} \left< \vec B (0) \cdot \vec A (t) \right> = 0 .
\end{equation}
More precisely, we observed that for finite time step $h$ there is a
slight systematic deviation ($\Delta (t) \ne 0$), which however tends
to zero for $h \to 0$. Furthermore, we found that $\Delta (t)$ quickly
decays to zero with increasing time $t$, and has both a positive
(small $t$) and a negative (larger $t$) contribution, such that
$\int_0^\infty dt \Delta (t)$ is very small. Such discretization
effects are the reason for our finding that the two functions
\begin{equation} \label{eq:definedoft}
D(t) \equiv D^{(K)} + D_1 (t) + D_2 (t)
\end{equation}
and
\begin{equation} \label{eq:definetildedoft}
\tilde D (t) \equiv D^{(K)} - D_1 (t) ,
\end{equation}
with
\begin{equation} \label{eq:defined1oft}
D_1 (t) = \frac{1}{3 N^2} \int_0^t d\tau
          \left< \vec A (0) \cdot \vec A (\tau) \right> 
\end{equation}
and
\begin{equation} \label{eq:defined2oft}
D_2 (t) = \frac{1}{3 N^2h^{1/2}} \int_0^t d\tau
          \left< \vec B (0) \cdot \vec A (\tau) \right> ,
\end{equation}
are slightly different, in particular for short times. This is seen in
Fig. \ref{fig:n6n25n50}, which shows $D(t)$ and $\tilde D(t)$ for $N =
6$, $25$, and $50$, as a function of $1/t$, with logarithmic abscissa.
This figure also demonstrates that our data are well--converged and
accurate enough to clearly discriminate between short-- and long--time
regimes.

Table \ref{table:d} summarizes our results for the diffusion
coefficient, where we list $D^{(K)}$, $D_1$ and $D_2$. The data
confirm the relation $2 D_1 + D_2 = 0$ within our error bars up to $N
= 50$. As it turned out that $D_2$ is much harder to sample than
$D_1$, we did not test the relation for the longer chains $N = 100$
and $200$, where our statistics is not sufficient. The data also show
that the relative contribution of the correction term systematically
increases with chain length (roughly $1 \%$ for $N = 6$, roughly $3.5
\%$ for $N = 200$).

\begin{figure}
 \begin{center}
 \epsfig{file=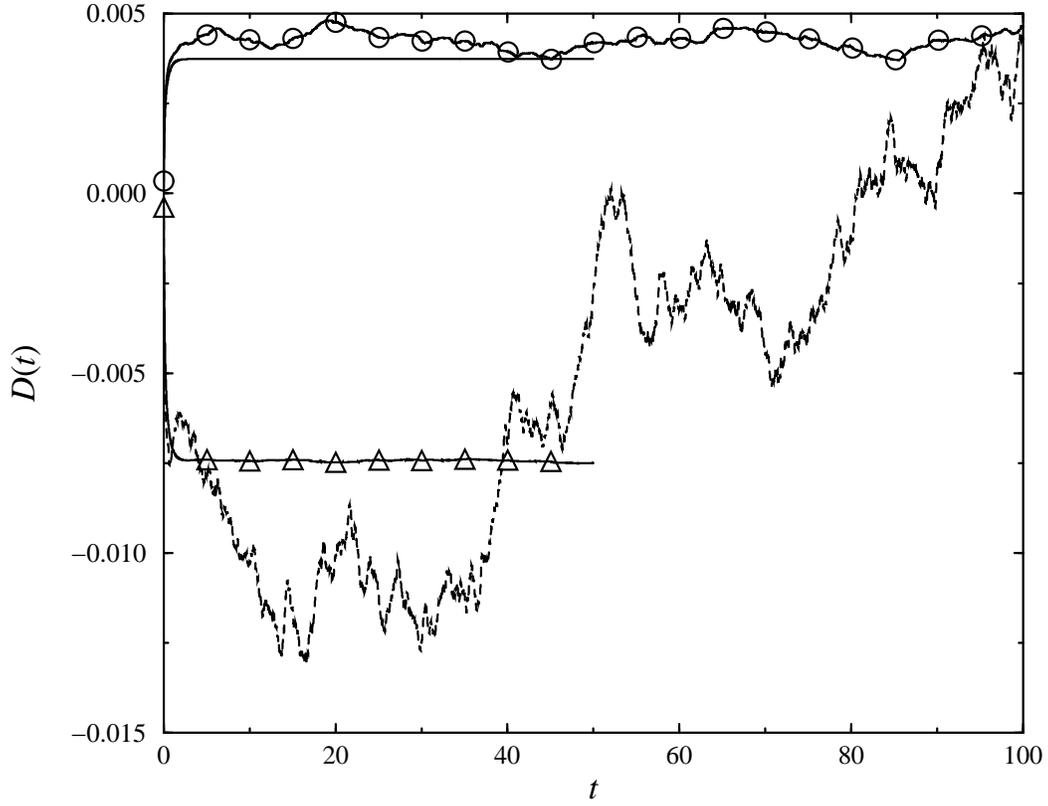,width=11cm,angle=-90}
 \end{center}
\caption{Time--dependent diffusion coefficients $D_1(t)$ and $D_2(t)$
  for $N = 6$ for different overall observation times: (i) Solid line:
  $D_1(t)$ (high resolution). (ii) Solid line with triangles: $D_2(t)$
  (high resolution). (iii) Solid line with circles: $D_1 (t)$ (low
  resolution). (iv) Dashed line: $D_2 (t)$ (low resolution). The parameters
  of the low--resolution calculation are adapted to those of Rey {\em et al.}
  \cite{rey:91}, as explained in the text.}
\label{fig:dtn6}
\end{figure}

Comparing our values for $D^{(K)}$ and $D_1$ with those of 
Ref. \onlinecite{rey:91}, we see that they are compatible
within error bars. However, Rey {\em et al.}
\cite{rey:91} have obtained values for $D$ which are larger than $D^{(K)}$.
In view of this puzzle, we have done a test run for $N = 6$, where we
increased the time step to their value $h = 0.01$, and decreased the
observation time to theirs ($0.2 \times 10^6$ time steps). Figure
\ref{fig:dtn6} shows our results for $D_1(t)$ and $D_2(t)$. One sees that the
statistical accuracy is sufficient to obtain an acceptable value for
$D_1$, but that it is by far not enough to estimate $D_2$. We believe
that this is the most likely explanation for the deviations observed
in Ref. \onlinecite{rey:91}.

\section{Scaling of the $\left< A A \right>$ Correlation Function}
\label{sec:scalinga2}

The systematic increase of the ratio $D_1 / D^{(K)}$ with chain length, as
seen from the data in Table \ref{table:d}, raises the question if that ratio
will saturate at a finite value, or keep on increasing, or maybe even tend to
zero for $N \to \infty$, after going through a maximum. We cannot give
a conclusive answer to this question; however, we have found some interesting
results concerning the issue. Rewriting the Green--Kubo formula for
$D_1$ as
\begin{equation} \label{eq:decomposed1}
D_1 = \frac{\left< A^2 \right>}{3 N^2} \int_0^\infty d t \, C_A (t) ,
\end{equation}
where $C_A$ is the normalized $A$--$A$ autocorrelation function,
\begin{equation}
C_A (t) = \frac{1}{\left< A^2 \right>}
          \left< \vec A (0) \cdot \vec A (t) \right> ,
\end{equation}
one sees that the $N$ dependence is clear if it is known for $\left<
A^2 \right>$ and for $\tau_A = \int_0^\infty d t \, C_A (t)$. As
$\left< A^2 \right>$ is a static average, let us discuss it first.

For {\em fluctuating} hydrodynamic interactions, we notice that
$\sum_j \tensor \mu_{ij} \cdot \vec F_j$ is nothing but the velocity
flow field generated at position of monomer $i$, due to all
the forces acting on the other monomers $j$. However, the system
is in thermal equilibrium. Therefore, one should expect that this
velocity is of order of a typical thermal velocity. Furthermore,
in equilibrium the flow velocities at different volume elements
are statistically uncorrelated. This picture suggests that
$\vec A$ is essentially the sum of $N$ statistically independent
random variables, each of which {\em does not} depend on $N$.
Therefore, the scaling
\begin{equation}
\left< A^2 \right> \propto N
\end{equation}
is expected from standard statistics, and this argument should be true
independently of the details of the chain conformations.

For {\em preaveraged} hydrodynamic interactions, however, this argument does
{\em not} hold (the preaveraging prevents $\tensor \mu_{ij}$ from
``thermalizing''). Here we rather write
\begin{equation}
\left< A^2 \right> = \sum_{ij} \sum_{kl}
\left< \mu_{ij}^{\alpha \beta} \right>
\left< \mu_{kl}^{\alpha \gamma} \right>
\left< F_j^\beta F_l^\gamma \right> ,
\end{equation}
where summation over repeated Cartesian indices, denoted by the Greek letters,
is implied. Exploiting the isotropy of the $\left< \vec F \otimes \vec F
\right>$ tensor (i.~e. its proportionality to the unit tensor) one finds
\begin{equation}
\left< A^2 \right> = \frac{1}{3}\sum_{ij} \sum_{kl}
\left< \mu_{ij}^{\alpha \beta} \right>
\left< \mu_{kl}^{\alpha \beta} \right>
\left< \vec F_j \cdot \vec F_l \right> .
\end{equation}
This is simple to study for the case of a {\em Gaussian} chain, since then the
$\left< \vec F \cdot \vec F \right>$ correlation is strictly short--ranged.
Indeed, for a random--walk chain, $\left< \vec F_i \cdot \vec F_j \right>$
must be zero if $i$ and $j$ are sufficiently far away from each other, since
in that case one can choose a ``pivot'' monomer between $i$ and $j$, and
rotate the ``right'' part of the chain around that monomer by a random angle,
without changing the statistical weight of the conformation. If $j$ is on the
rotated part, $\vec F_j$ is changed, while $\vec F_i$ is unchanged. Thus one
shows $\left< \vec F_i \cdot \vec F_j \right> = - \left< \vec F_i \cdot \vec
  F_j \right> = 0$.  This argument holds whenever it is possible to find a
pivot monomer, i.~e. for $\left\vert i - j \right\vert \ge 2$. Thus the only
remaining correlations are those for $i = j$ and $\left\vert i - j \right\vert
= 1$, in which case the correlation is obvious, due to the spring interaction
with the neighboring monomer. In case of an excluded--volume chain we rather
expect a power--law decay \cite{deGennes:79}, related to the probability of
loops of length $\left\vert i - j \right\vert$. In what follows, we will
therefore, for simplicity, focus on the Gaussian case. Noticing $\left<
  \tensor \mu_{ij} \right> \propto \left\vert i - j \right\vert^{-1/2}$, we
thus find
\begin{equation}
\left< A^2 \right> \propto \int_0^N dx \int_0^N dy \int_0^N dz
\left\vert x - y \right\vert^{-1/2}
\left\vert z - y \right\vert^{-1/2}
\end{equation}
(the short range of $\left< \vec F \cdot \vec F \right>$ reduces
the number of integrations from four to three). A trivial transformation
to reduced variables $x / N$ etc. then shows
\begin{equation}
\left< A^2 \right> \propto N^2
\end{equation}
for preaveraged hydrodynamics in the Gaussian case.

\begin{figure}
 \begin{center}
 \epsfig{file=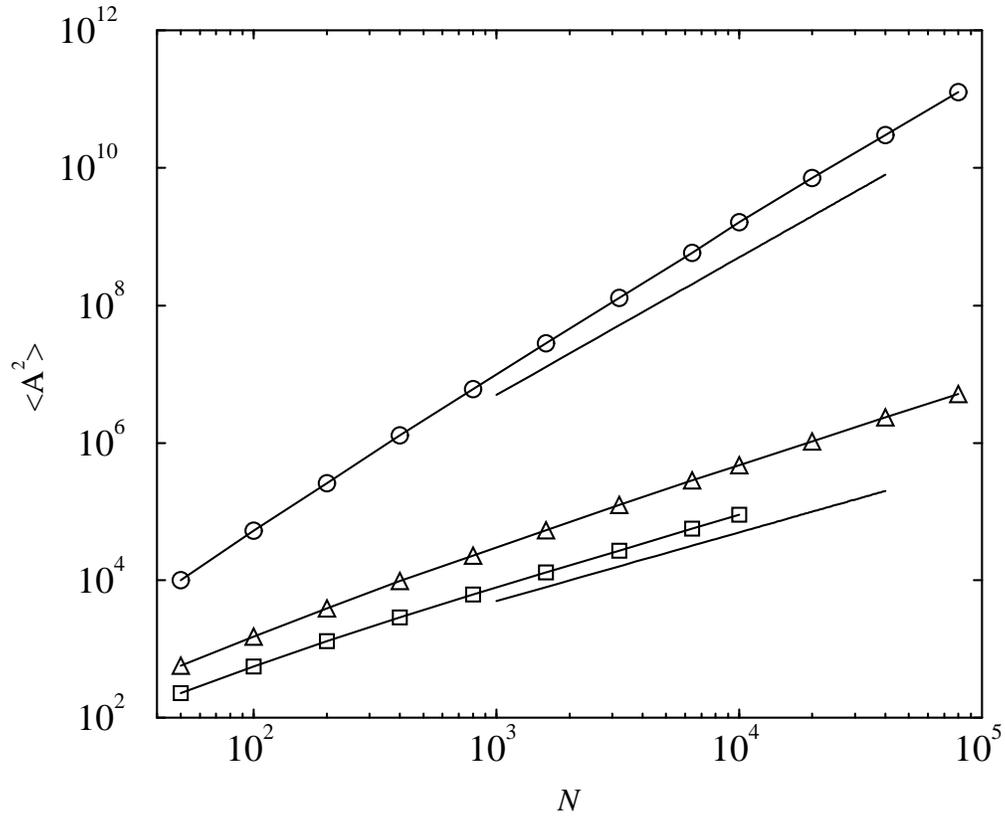,width=11cm,angle=-90}
 \end{center}
\caption{Static average $\left\langle A^{2}\right\rangle$ as a function of
  chain length $N$ for different cases: (i) Gaussian chain with preaveraged
  hydrodynamics (circles); (ii) Gaussian chain with fluctuating hydrodynamics
  (triangles); (iii) excluded--volume chain with fluctuating hydrodynamics
  (squares). The slopes of the solid lines indicate the power laws $N^2$ and
  $N^1$.}
\label{fig:a2}
\end{figure}

We have tested these predictions numerically, and exploited the fact that
$\left< A^2 \right>$ is a static average, and, as such, amenable to efficient
Monte Carlo procedures. This is particularly true for the Gaussian case, where
one simply generates a sample of chains. We were therefore able to study this
case up to chains of length $N = 0.8 \times 10^5$. However, we restricted
ourselves, for simplicity, to Oseen--like hydrodynamic interactions, where we
studied both the fluctuating and the preaveraged case. Apart from this, we
also studied the behavior for our model (fluctuating hydrodynamics,
excluded--volume chains) up to chains of length $N = 10^4$. In this case, we
used the full RPY interaction, and generated the conformations by a
combination of the pivot algorithm \cite{lal:69,stellman:72} with local moves.
For every chain length, $0.2 \times 10^5$ pivot moves, and $100$ times as many
local moves, were used. The results are presented in Fig. \ref{fig:a2}; indeed
reasonable agreement with our predictions is found.

The scaling laws for $\left< A^2 \right>$ have an interesting implication for
the dynamics of $\vec A$. Writing $\left< A^2 \right> \propto N^x$ where
$x = 1, 2$ for the discussed cases, and $\tau_A \propto N^y$, we find
from Eq. \ref{eq:decomposed1}
\begin{equation}
D_1 \propto N^{x + y - 2} .
\end{equation}
On the other hand, it is well--established that $D^{(K)}$ is proportional to
$N^{- \nu}$ where $\nu$ is $1/2$ for Gaussian chains, and $0.59$ for
excluded--volume chains. It also strongly believed that this is the asymptotic
scaling law for $D$. This, however, implies that $D_1$ must decay sufficiently
quickly as a function of $N$ --- otherwise $D_1$ would ultimately dominate and
spoil the scaling of $D$. More precisely, one expects $D_1 \propto N^{-\phi}$
with $\phi \ge \nu$. Combined with the previous consideration, this yields
$\phi = 2 - x - y \ge \nu$ or $y \le 2 - x - \nu$, i.~e. $y \le 1 - \nu$ for
fluctuating hydrodynamics, and $y \le - \nu$ for preaveraged hydrodynamics of
a Gaussian chain. This is a quite counter--intuitive result, since it implies
that $\tau_A$ would increase only very weakly with chain length for
fluctuating hydrodynamics, while it would even {\em decrease} for preaveraged
hydrodynamics! Naively, one would rather expect that $\vec A$, as a collective
quantity, decays on the same time scale as the overall polymer conformations,
i.~e. $\tau_A \propto \tau_Z \propto N^{3 \nu}$ (this latter relation is the
standard Zimm scaling law \cite{doi:86}, and implies a rather sharp increase
with $N$). We have thus found a ``collective'' quantity which apparently
decays much more rapidly than the chain as a whole. We believe this issue
deserves further attention; in particular, we think it would be very desirable
to try to understand the underlying physical mechanisms governing the
relaxation of $\vec A$ somewhat better.

\begin{figure}
 \begin{center} 
 \epsfig{file=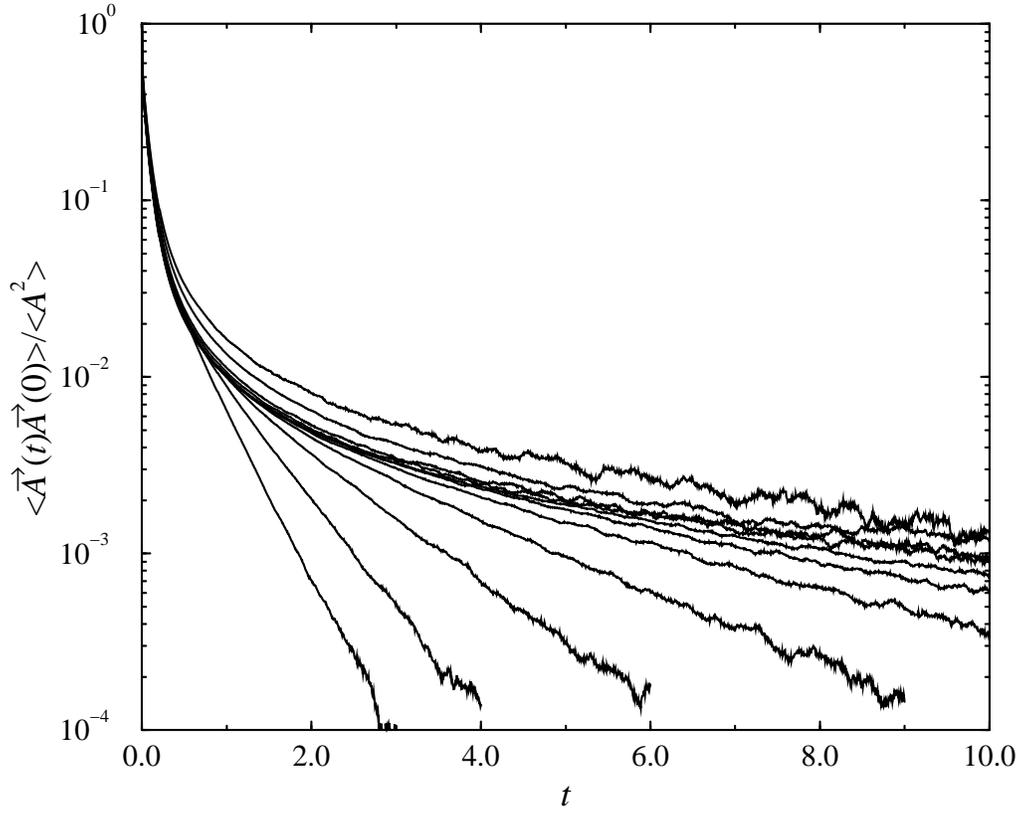,width=11cm,angle=-90}
 \end{center} 
 \caption{Normalized autocorrelation function $\left\langle \vec
   A(t) \vec A(0)\right\rangle / \left\langle \vec A^2 \right\rangle$,
   for $N = 6$, 8, 11, 15, 20, 25, 30, 40,
   50, 100, 200. The chain length increases from left to right.}
\label{fig:aa}
\end{figure}

Our numerical data from the BD simulation (i.~e. for fluctuating
hydrodynamics, and excluded--volume chains) can only give us very limited
hints on the behavior of $\tau_A$ as a function of $N$, since, due to the
overall computational demand, we were not able to simulate the dynamics with
sufficient accuracy for chains longer than $N = 200$. Our data for $C_A (t)$
are presented in Fig. \ref{fig:aa}. Apparently the correlation function has
two distinct time regimes. In the short--time regime ($t < t_{0}$), the curves
are practically superimposable. This is similar to the observations made by
Fixman \cite{fixman:83a}. In the long--time regime ($t > t_{0}$), the
correlation functions decay exponentially with a correlation time $\tau_D$,
$C_A (t) \propto \exp( - t / \tau_D )$. Figure 6 shows our data for
$\tau_D$.  Indeed $\tau_D$ increases with chain length; however, the
observed behavior in our limited $N$ window is anything but a power law.
According to our previous considerations, the increase of $\tau_D$ should not
be stronger than $N^{1 - \nu}$. Indeed this condition seems to be
satisfied in the regime of longer chains.

\section{Further Results}
\label{sec:results}

\subsection{Static Scaling Properties}

The radius of gyration and end--to--end distance are given by
\begin{equation}
\left\langle R_{g}^{2} \right\rangle = 
\frac{1}{2N^2}\sum_{ij}\left\langle r_{ij}^{2} \right\rangle
\end{equation}
and
\begin{equation}
\left\langle R_{e}^{2}\right\rangle=\left\langle \left(\vec r_{N}-\vec 
    r_{1}\right)^{2} \right\rangle .
\end{equation}
The theoretical scaling for these static properties is
\begin{equation}
\left\langle R_{g}^{2}\right\rangle \propto\left\langle R_{e}^{2} 
\right\rangle \propto \left( N-1 \right)^{2\nu} .
\end{equation}
In good solvent, the scaling exponent has the theoretical value of $\nu
\approx 0.588$ from renormalization group calculations and Monte Carlo
simulations \cite{li:95}. The log--log fits of $\left\langle R_{e}^{2}
\right\rangle$ and $\left\langle R_{g}^{2}\right\rangle$ vs. $N-1$ yield
the exponents $2\nu=1.187 \pm 0.003$ and $2\nu=1.133 \pm 0.006$, respectively
(see Fig. \ref{fig:rgren}), which is similar to the results by Rey {\em et
  al.} \cite{rey:91}.

Similarly, the static structure factor
\begin{eqnarray}
S(k) & = & \frac{1}{N} \sum_{ij} 
           \left\langle \exp (i\vec k \cdot \vec r_{ij}) \right\rangle
     \nonumber \\
     & = & \frac{1}{N} \sum_{ij}
           \left\langle \frac{\sin (kr_{ij})}{kr_{ij}} \right\rangle,
\end{eqnarray}
which is measured in scattering experiments, obeys the scaling relation
\begin{equation}
S(k) \propto k^{-1/\nu}
\end{equation}
in the regime $R_{g}^{-1} \ll k \ll b^{-1}$. By fitting a power law to our
data we get the value (see Fig. \ref{fig:struct}) $\nu=0.575 \pm 0.004$.

We have also obtained the first cumulant (or initial decay rate), $\Omega
(k)$, of the dynamic structure factor $S(k,t)$, defined as
\begin{equation} 
\Omega(k) = - \lim_{t\to 0} \frac{d}{dt} 
\left( \frac{S(k,t)}{S(k,0)} \right) .
\end{equation}
Akcasu {\em et al}. \cite{akcasu:80,benmouna:80} have shown that $\Omega (k)$
can be written as
\begin{equation} \label{eq:qels}
\Omega (k) = \frac{\sum_{ij} \left\langle 
\vec k \cdot \tensor D_{ij} \cdot \vec k 
\exp ( i\vec k \cdot \vec r_{ij} ) \right\rangle}
{\sum_{ij} \left\langle \exp ( i\vec k \cdot \vec r_{ij}) \right\rangle}.
\end{equation}
The orientational averaging in Eq. (\ref{eq:qels}) is easily done for the RPY
tensor \cite{rey:91,burchard:80}. For the denominator one obtains
\begin{equation} \label{eq:qels1}
\left\langle \exp ( i\vec k \cdot \vec r_{ij} ) \right\rangle_{or} = 
\frac{\sin z}{z}
\end{equation}
with $z = k r_{ij}$.
In the numerator we find for $i = j$
\begin{equation}
\left\langle( \vec k \cdot \tensor D_{ij} \cdot \vec k )
 \exp ( i\vec k \cdot \vec r_{ij} ) \right\rangle_{or} =
\frac{k_B T}{6 \pi \eta a} k^2 .
\end{equation}
For $i \ne j$ one obtains instead
\begin{eqnarray}
&   & \left\langle( \vec k \cdot \tensor D_{ij} \cdot \vec k )
      \exp ( i\vec k \cdot \vec r_{ij} ) \right\rangle_{or} 
      \nonumber \\
& = & \frac{k_B T}{4 \pi \eta r_{ij}} k^2 
      \Bigg[ \left( 1 - \frac{2}{3} \frac{a^2}{r_{ij}^2} \right)
      \frac{\sin z}{z} \nonumber \\
&   & + \left( 1 - 2 \frac{a^2}{r_{ij}^2} \right)
      \left( \frac{\cos z}{z^2} - \frac{\sin z}{z^3} \right) \Bigg]
\end{eqnarray}
in the case of large distances $r_{ij} \ge 2 a$, while
\begin{eqnarray}
&   & \left\langle( \vec k \cdot \tensor D_{ij} \cdot \vec k )
      \exp ( i\vec k \cdot \vec r_{ij} ) \right\rangle_{or}
      \nonumber \\
& = & \frac{k_B T}{6 \pi \eta a} k^2
      \Bigg[ \left( 1 - \frac{3}{16} \frac{r_{ij}}{a} \right) \frac{\sin z}{z}
      \nonumber \\
&   & + \frac{3}{16} \frac{r_{ij}}{a} 
      \left( \frac{\cos z}{z^2} - \frac{\sin z}{z^3} \right) \Bigg]
\end{eqnarray}
for $r_{ij} < 2 a$. We should mention that there are some typographical errors
both in Ref.\onlinecite{rey:91} and Ref.\onlinecite{burchard:80}. The
right--hand term of Eq. (\ref{eq:qels}) is therefore directly calculated from
the trajectories. In the $k\to 0$ limit, $\Omega (k)$ reflects exclusively the
translational motion contribution to the chain dynamics. Therefore, the
Kirkwood formula can be recovered from the first cumulant as
\begin{equation} \label{eq:dk1}
D^{(K)}=\lim_{k\to 0}\Omega(k)/k^{2}.
\end{equation}
We have obtained $D^{(K)}$ through Eq. (\ref{eq:dk1}) from the intercept of a
fitting of $\Omega(k)/k^{2}$ vs. $k$ in the $k\to 0$ limit shown in Fig.
\ref{fig:qels}. These values are exactly the same as those obtained from Eq.
(\ref{eq:dk}), which constitutes a further verification of the consistency of
our numerical method. {From} Fig. \ref{fig:qels}, a universal dependence of the
type $\Omega(k)/k^{2} \propto k$ is also obtained in the scaling regime
$R_{g}^{-1} \ll k \ll b^{-1}$.

\subsection{Dynamic Scaling Properties}

An approximately exponential behavior of the time--correlation function
$\left\langle \vec{R}_{e}(t) \cdot \vec{R}_{e}(0) \right\rangle$, where
$\vec{R}_{e}$ is the end--to--end vector, is observed (see Fig.
\ref{fig:endt}). We have extracted the relaxation times (Zimm times)
$\tau_{Z}$ corresponding to this behavior. $\tau_{Z}$ is related
to the orientational diffusion of the end--to--end vector. A log--log
fit of $\tau_{Z}$ vs. $N$ yields a slope of $1.71 \pm 0.01$
(Fig. \ref{fig:endtN}), which is close to the theoretial value $3\nu$
with hydrodynamic interactions.

The dynamic structure factor 
\begin{equation}
S(k,t) = \frac{1}{N}
\sum_{i j} \left< \exp\left[ i \vec k \cdot 
\left( \vec r_i(t) - \vec r_j (0) \right) \right] \right> 
\end{equation}
is predicted to exhibit scaling behavior \cite{doi:86} if both wave number and
time are in the scaling regime, i.~e.  $R_{g}^{-1} \ll k \ll b^{-1}$ and
$\tau_{0} \ll t \ll \tau_{Z}$, where $\tau_{0}$ is the microscopic time and
$\tau_{Z}$ is the Zimm time, the longest relaxation time of the chain. Fig.
\ref{fig:ds} gives a nice data collapse for the expected form
\begin{equation}
\frac{S(k,t)}{S(k,0)} = f \left( k^2 t^{2/3} \right)
\end{equation}
in log--linear represention. 

These scaling results demonstrate the internal consistency of our simulation,
and in all cases agreement with the pertinent theories and experimental
results.

\section{Summary}
\label{sec:summary}

The present study has shown that Brownian Dynamics simulations are
able to attack the problem of translational diffusion of polymer
chains with hydrodynamic interaction and excluded volume. It has also
highlighted the necessity of substantial statistical effort in order
to obtain reliable data.  While the standard picture of static and
dynamic scaling is reproduced, as in previous studies, the novel
aspect is the calculation of the diffusion coefficient to sufficiently
high accuracy, such that the difference between the short--time
Kirkwood value and the asymptotic long--time value could be resolved
unambiguously. The numerical data are in perfect agreement with the
theoretical predictions, both concerning the short--time value, and
the crossover to the long--time value described by Fixman's
Green--Kubo formula. It turns out that the long--time value is a few
percent less than the short--time value. For Fixman's Green--Kubo
integrand we find two remarkable results, namely that its initial
value behaves very differently for preaveraged vs. fluctuating
hydrodynamics, and that the correlation function, though describing a
global property of the chain, must decay substantially faster than the
conformations, in order to avoid a violation of dynamic scaling. Our
numerical data are in reasonable agreement with these considerations,
but not fully conclusive since only short chains were accessible. More
work on this issue, in particular aimed at a better physical
understanding, is clearly desirable.


\bibliographystyle{aip}
\bibliography{jcpliu}

\end{multicols}

\clearpage

\widetext


\begin{table}
\begin{tabular}{cccccc} 
   
$N$ & $D^{(K)}$           & $D_{1}$            & $D_{2}$           &$2D_{1} +
D_{2} $  & $\left\vert D_{1} + D_{2} \right\vert / D^{(K)}$ \\ \hline
 6  & $0.3544\pm10^{-4} $   & $ 0.00408\pm10^{-5} $  & $
 -0.0081\pm10^{-4} $ & $ 0.0000\pm10^{-4}$  & $ 0.0113\pm3\times10^{-4}$\\
 8  & $0.3011\pm10^{-4} $   & $ 0.00441\pm10^{-5} $  & $
 -0.0087\pm10^{-4} $ & $ 0.0001\pm10^{-4}$  & $ 0.0143\pm3\times10^{-4}$\\
11  & $0.2513\pm10^{-4} $   & $ 0.00450\pm10^{-5} $  & $
-0.0089\pm10^{-4} $ & $ 0.0001\pm10^{-4} $  & $ 0.0175\pm4\times10^{-4}$\\
15  & $0.2106\pm10^{-4} $   & $ 0.00443\pm10^{-5} $  & $
-0.0088\pm10^{-4} $ & $ 0.0000\pm10^{-4} $  & $ 0.0214\pm5\times10^{-4}$\\
20  & $0.1788\pm10^{-4} $   & $ 0.00443\pm10^{-5} $  & $
-0.0084\pm10^{-4} $ & $ 0.0001\pm10^{-4} $  & $ 0.0234\pm6\times10^{-4}$\\
25  & $0.1573\pm10^{-4} $   & $ 0.00422\pm10^{-5} $  & $
-0.0079\pm10^{-4} $ & $ 0.0000\pm10^{-4} $  & $ 0.0249\pm6\times10^{-4}$\\
30  & $0.1416\pm10^{-4} $   & $ 0.00399\pm10^{-5} $  & $
-0.0076\pm10^{-4} $ & $ 0.0001\pm10^{-4} $  & $ 0.0267\pm7\times10^{-4}$\\
35  & $0.1298\pm10^{-4} $   & $ 0.00381\pm10^{-5} $  & $
-0.0072\pm10^{-4} $ & $ 0.0000\pm10^{-4} $  & $ 0.0275\pm8\times10^{-4}$\\
40  & $0.1201\pm10^{-4} $   & $ 0.00363\pm10^{-5} $  & $
-0.0069\pm10^{-4} $ & $ 0.0000\pm10^{-4} $  & $ 0.0287\pm8\times10^{-4}$ \\
45  & $0.1123\pm10^{-4} $   & $ 0.00345\pm10^{-5} $  & $
-0.0066\pm10^{-4} $ & $ 0.0000\pm10^{-4} $  & $ 0.0292\pm9\times10^{-4}$\\
50  & $0.1055\pm10^{-4} $   & $ 0.00332\pm10^{-5} $  & $
-0.0063\pm10^{-4} $ & $ 0.0000\pm10^{-4} $  & $
0.0298\pm9\times10^{-4}$ \\
100 & $0.0718\pm10^{-4} $   & $ 0.0022\pm10^{-4} $  & $ $ & $  $  & $
0.031\pm1\times10^{-3}$ \\
200 & $0.0428\pm10^{-4} $   & $ 0.0015\pm10^{-4} $  & $ $ & $  $  & $
0.035\pm1\times10^{-3}$ \\
\end{tabular}
\caption{The diffusion coefficients $D^{(K)}$, $D_{1}$, $D_{2}$, as
  well as $2 D_{1} + D_{2}$, and $\left\vert D_{1} + D_{2}\right\vert / 
  D^{(K)}$, as defined in the text, for different chain lengths $N$.
  Note that $D_2$ was not sampled for $N = 100$ and $200$, for
  reasons of poor statistics, and that hence for these chains we
  have assumed $\left\vert D_{1} + D_{2}\right\vert = D_1$. }
\label{table:d}
\end{table}


\clearpage

\clearpage

\clearpage

\clearpage

\clearpage

\clearpage

\begin{figure}
 \begin{center}
 \epsfig{file=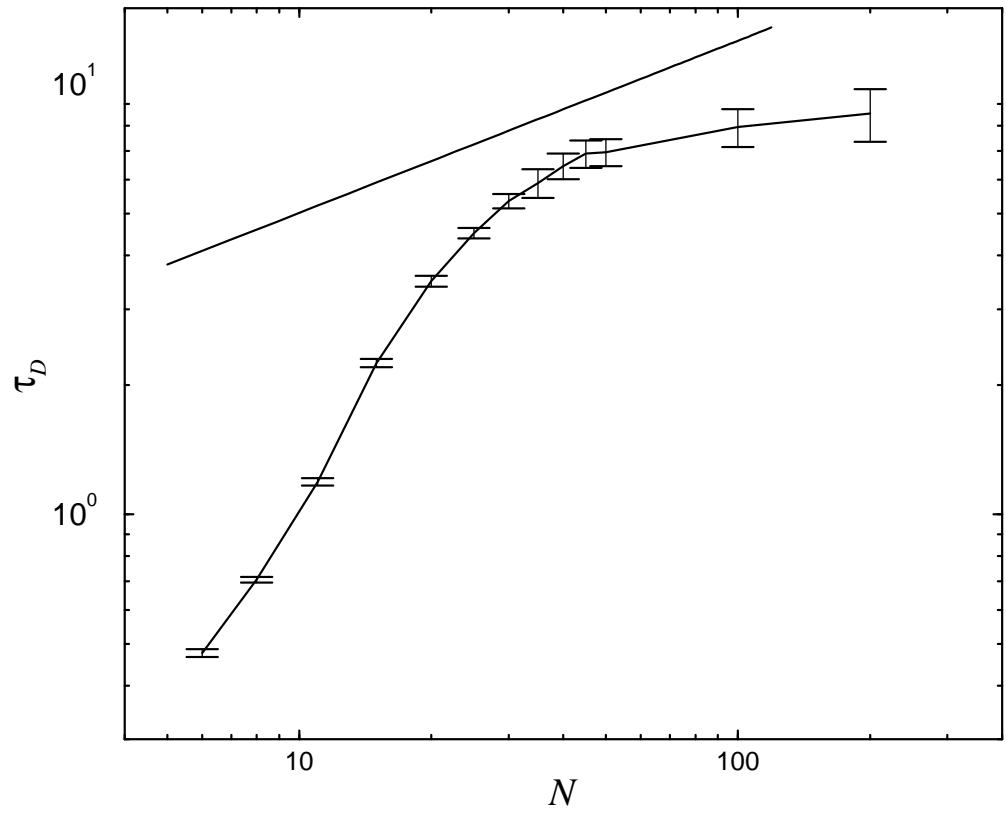,width=11cm,angle=-90}
 \end{center}
 \caption{Chain length scaling for $\tau_{D}$. The slope of the solid line
    indicates the power law $N^{1-\nu}$.}
\label{fig:tau}
\end{figure}

\clearpage

\begin{figure}
\begin{center}
\epsfig{file=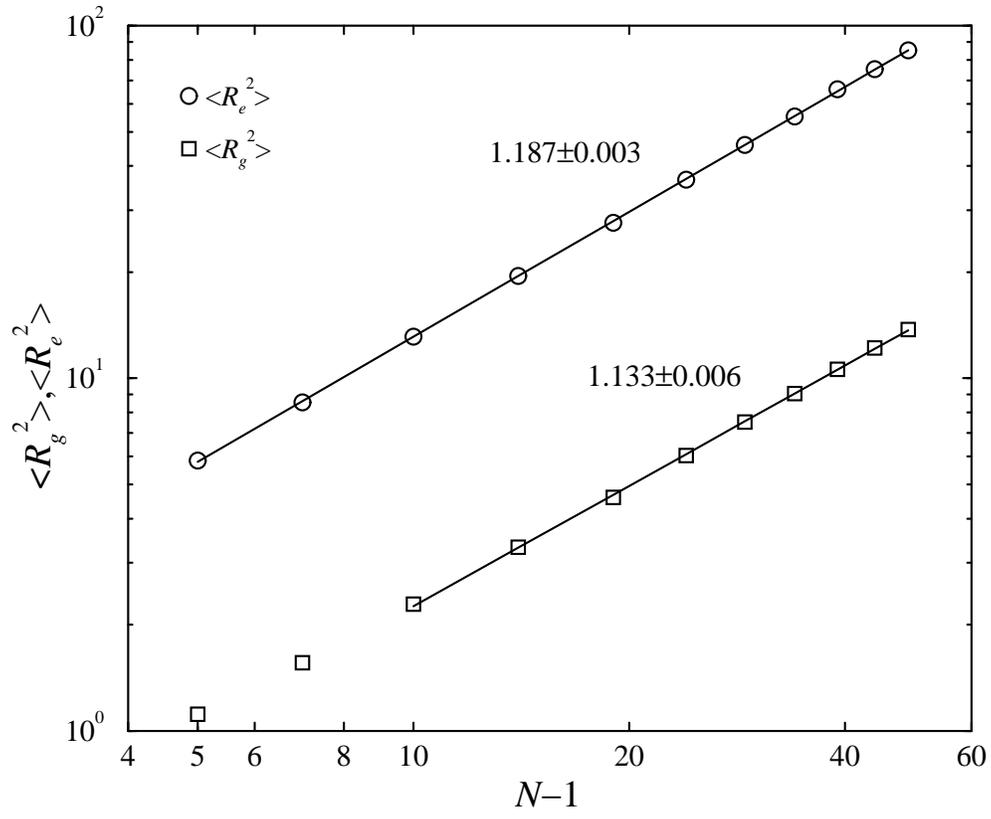,width=11cm,angle=-90}
\end{center}
\caption{Chain length scaling for $\left\langle R_{g}^{2}
  \right\rangle$ and $\left\langle R_{e}^{2}\right\rangle$. The error
  bars are smaller than the symbol size.}
\label{fig:rgren}
\end{figure}

\clearpage

\begin{figure}
 \begin{center}
 \epsfig{file=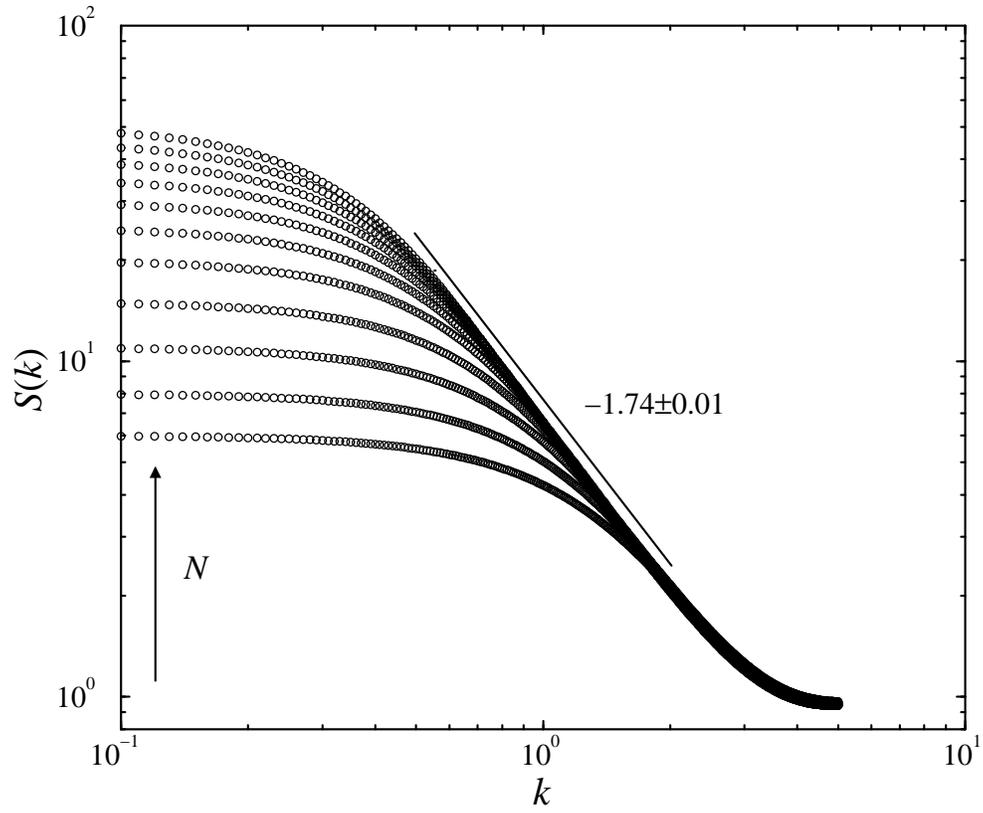,width=11cm,angle=-90}
 \end{center}
\caption{The static structure factor of the chain, for $N=6$, 8, 11,
  15, 20, 25, 30, 35, 40, 45, 50. The chain length increases along
  the arrow.}
\label{fig:struct}
\end{figure}

\clearpage

\begin{figure}
 \begin{center}
 \epsfig{file=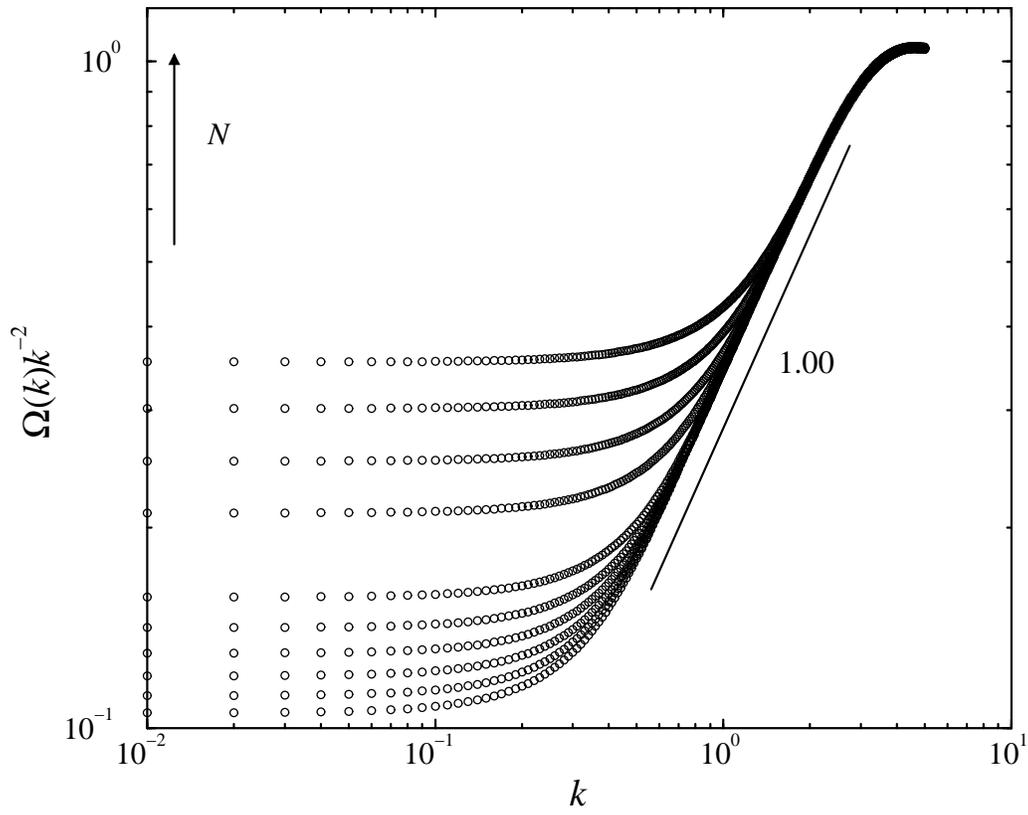,width=11cm,angle=-90}
 \end{center}
\caption{$\Omega(k)/k^{2}$ vs. $k$ for $N=6$, 8, 11, 15, 20, 25, 30,
  35, 40, 45, 50. The chain length increases along the arrow.}
\label{fig:qels}
\end{figure}

\clearpage

\begin{figure}
 \begin{center}
 \epsfig{file=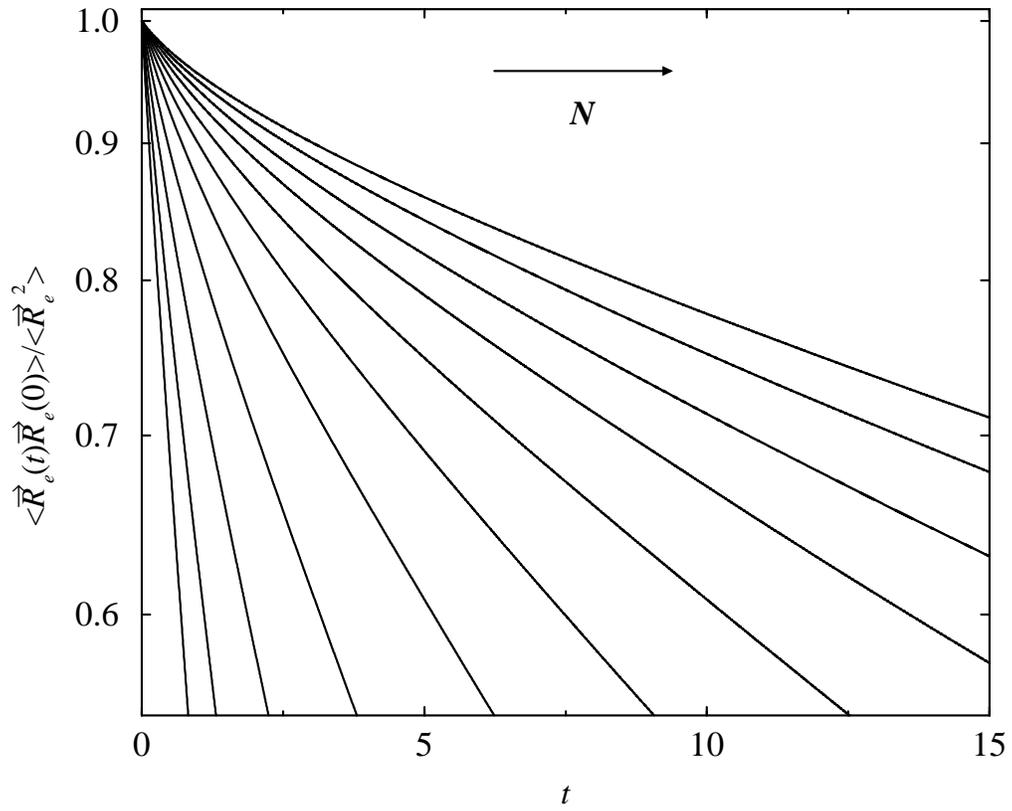,width=11cm,angle=-90}
 \end{center}
\caption{Normalized autocorrelation function of the end-to-end vector $\vec
  {R}_{e}$ for $N=6$, 8, 11, 15, 20, 25, 30, 35, 40, 45, 50. The chain length
  increases along the arrow.}
\label{fig:endt}
\end{figure}

\clearpage

\begin{figure}
 \begin{center}
 \epsfig{file=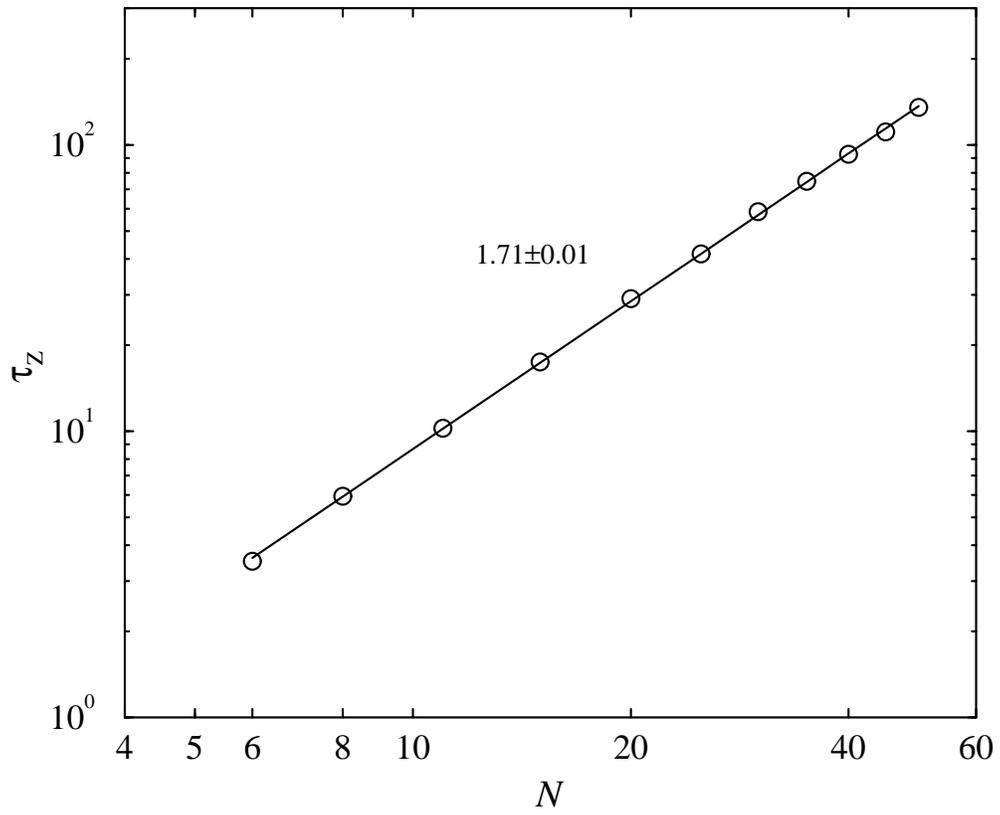,width=11cm,angle=-90}
 \end{center}
\caption{Chain length scaling for $\tau_{Z}$ obtained from the
  autocorrelation function of the end--to-end vector.
  The error bars are smaller than the symbol size.}
\label{fig:endtN}
\end{figure}

\clearpage

\begin{figure}
 \begin{center}
 \epsfig{file=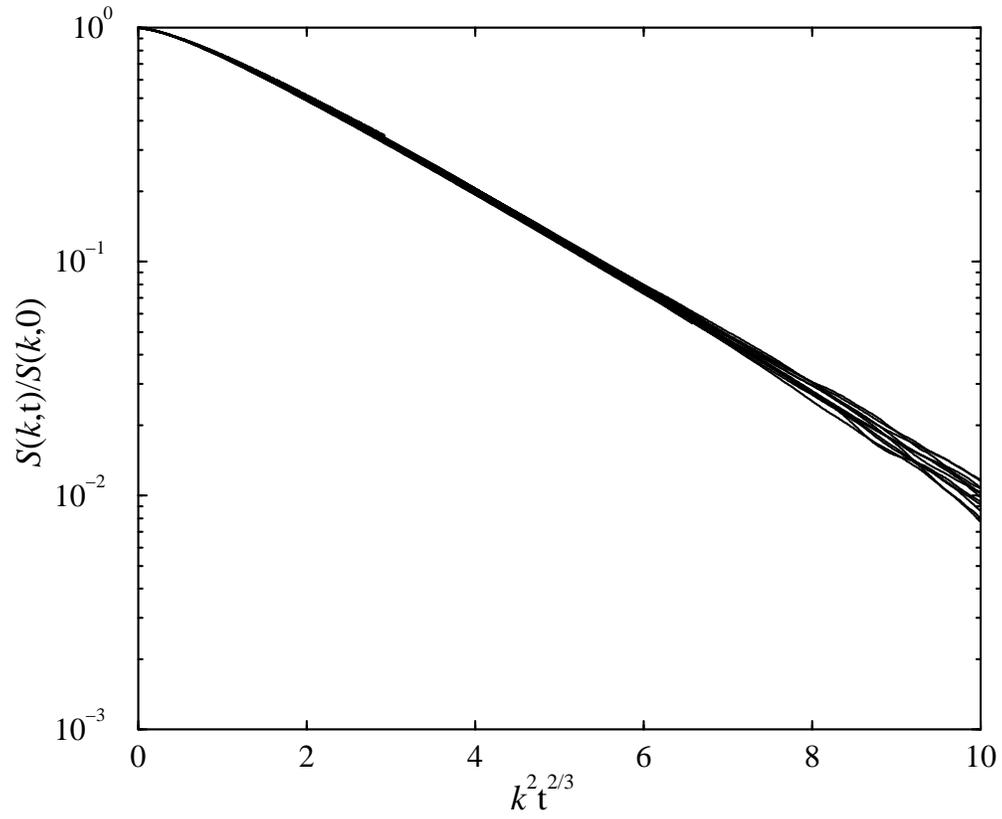,width=11cm,angle=-90}
 \end{center}
\caption{Log--linear scaling plot of the decay of the dynamic structure 
  factor for $N=15$, 20, 25, 30, 35, 40, 45, 50 for Zimm scaling:
  $S(k,t)/S(k,0)$ vs. $k^{2}t^{2/3}$. We restricted the wave numbers
  to the values $k=1.0$, 1.5, 2.0, and the time regime to
  $5 \le t \le 20$. }
\label{fig:ds}
\end{figure}

\end{document}